\documentclass[onecolumn,showpacs,superscriptaddress,preprintnumbers,nofootinbib]{revtex4}

\usepackage{bm}
\usepackage{feynmp}
\usepackage{simplewick}
\usepackage{graphics,graphicx}
\usepackage{amsmath,amssymb}
\usepackage{color}
\usepackage{booktabs,multirow}

\newcommand{\Bscal}{B^*_0}

\newcommand{\bea}{\begin{eqnarray}}
\newcommand{\eea}{\end{eqnarray}}
\newcommand{\GeV}{\mathrm{GeV}}
\newcommand{\beq}{\begin{equation}}
\newcommand{\eeq}{\end{equation}}
\newcommand{\GEVP}{\mathrm{GEVP}}

\makeatother

\begin{document}
\preprint{\tt LPT Orsay 14-74}
\preprint{\tt DAMTP-2014-56}
\vspace*{22mm}

\title{Pion couplings to the scalar $B$ meson}

\author{Beno\^it~Blossier}
\affiliation{Laboratoire de Physique Th\'eorique\footnote[1]{Unit\'e Mixte de Recherche 8627 du Centre National de la Recherche Scientifique}, CNRS et Universit\'e Paris-Sud XI,  B\^atiment 210, 
91405 Orsay Cedex, France}
\author{Nicolas~Garron\footnote[2]{On leave from School of Mathematics, Trinity College, Dublin 2, Ireland}}
\affiliation{Department of Applied Mathematics \& Theoretical Physics, University of Cambridge,
Wilberforce Road, Cambridge CB3 0WA, United Kingdom}
\author{Antoine~G\'erardin}
\affiliation{Laboratoire de Physique Th\'eorique\footnote[1]{Unit\'e Mixte de Recherche 8627 du Centre National de la Recherche Scientifique}, CNRS et Universit\'e Paris-Sud XI,  B\^atiment 210, 
91405 Orsay Cedex, France}
\affiliation{Laboratoire de Physique Corpusculaire de Clermont-Ferrand\footnote[3]{Unit\'e Mixte de Recherche 6533 CNRS/IN2P3 -- Universit\'e Blaise Pascal}, Campus des C\'ezeaux,
24 avenue des Landais, BP 80026, 63171 Aubi\`ere Cedex, France}

\begin{abstract}
We present two-flavor lattice QCD estimates of the hadronic couplings $g_{B^*_0 B \pi}$ and $g_{B_1^* B_0^* \pi}$ 
that parametrise the non leptonic decays $B^*_0 \to B \pi$ and $B^*_1 \to B_0^* \pi$. We use CLS two-flavour gauge ensembles.
Our framework is the Heavy Quark Effective Theory 
(HQET) in the static limit and solving a Generalized Eigenvalue Problem (GEVP) reveals crucial to disentangle the
$B^*_0$($B^*_1$) state from the $B \pi$($B^*\pi$) state. 
This work brings us some experience on how to treat the possible 
contribution from multihadronic states to correlation functions calculated on the lattice, especially when $S$-wave states are involved. 

\end{abstract}

\maketitle

\section{\label{Introduction}Introduction}

\begin{figure}[t]
	\begin{minipage}[c]{0.49\linewidth}
	\centering 
	\includegraphics*[width=4.5cm]{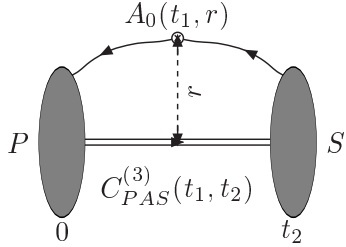}
	\end{minipage}
	\caption{Three-point correlation function used by \cite{BecirevicZZA} to compute $A_+(\Delta^2=q^2)$.}	
\label{figA0}
\end{figure}

Heavy Meson Chiral Perturbation Theory (HM$\chi$PT) \cite{WiseHN, YanGZ} is commonly used to extrapolate lattice data in the heavy-light sector to the physical point. Relying on Heavy Quark Symmetry and the (spontaneously broken) chiral 
symmetry, an effective Lagrangian is derived where heavy-light mesons fields \cite{FalkYZ} couple to a Goldstone field
via derivative operators. In the static limit, the total angular momentum of the light degrees of freedom, $\vec{j}_l=\vec{s}_l+\vec{L}$, is conserved independently of the total angular momentum $J=j_l \pm 1/2$. The pseudoscalar ($B$) and the vector ($B^*$) mesons belong to the doublet $j_l^P=(1/2)^-$ corresponding to $L=0$ whereas the scalar ($B_0^*$) and the axial ($B_1^*$) mesons belong to the positive parity doublet $j_l^{P}=(1/2)^+$ corresponding to $L=1$ (see Table~\ref{tab:spectrum}). Equivalently to the low energy constants that parametrize the well known chiral Lagrangian, hadronic couplings enter the effective theory under discussion, that is 
particularly suitable to describe processes with emission of soft pions, i.e. $H_1(J_1) \to H_2(J_2) \pi$ where $H_i$ is a heavy-light meson, 
and $p_\pi \ll \Lambda_\chi \sim 1$ GeV. The associated pionic couplings are $g_{H_1(J_1)H_2(J_2)\pi}$ and they cannot be computed in perturbation theory. When the negative $j^P=(1/2)^{-}$ and positive $j^P=(1/2)^+$ parity states are taken into account, the effective Lagrangian is parametrized by three couplings $\widehat{g}$, $\widetilde{g}$ and $h$. The first coupling, $\widehat{g}$, relates transitions between mesons belonging to the same doublet $J^P=(1/2)^{-}$ and has been precisely measured on the lattice \cite{OhkiPY} - \cite{BulavaEJ}. On the contrary, the last two couplings are less precisely known.
The residue at the poles of form factors in heavy to light semileptonic decays \cite{DescotesGenonHH} is also expressed in terms of those couplings. In that respect, the channel $B^{*}_0 \to B \pi$ is very interesting: 
\renewcommand{\arraystretch}{1.15}
\setlength{\tabcolsep}{0.2cm}
\begin{table}[b]
\begin{center}
\begin{tabular}{|c|c|c|c|}
\hline 
$L$				& 	$j^P_l$					&  $J^P$	 	&  	state		\\ 
\hline
\multirow{2}{*}{$0$}	&	\multirow{2}{1cm}{$(1/2)^{-}$}	&	$0^{-}$	&	$B^{\ }$	\\
				&							&	$1^{-}$	&	$B^{*}$   	\\ 
\hline
\multirow{2}{*}{$1$}	&	\multirow{2}{1cm}{$(1/2)^{+}$}	&	$0^{+}$	&	$B_0^{*}$	\\
				&							&	$1^{+}$	&	$B_1^{*}$	\\ 
\hline
\end{tabular}
\end{center}
\caption{Quantum numbers of the ground state $B$ meson and its first orbital excitations.}
\label{tab:spectrum}
\end{table}
\beq\nonumber
\Gamma(B^{*0}_0 \to B^+ \pi^-)=\frac{1}{8\pi} \, g^2_{B^*_0 B \pi}  \,
\frac{|\vec{q}_\pi|}{m^2_{B^*_0}} \,,\quad 
|\vec{q}_\pi|=\frac{\sqrt{[m^2_{\Bscal}-(m_B+m_\pi)^2][m^2_{\Bscal}-(m_B-m_\pi)^2]}}{2m_{\Bscal}} \,. 
\eeq
The HM$\chi$PT Lagrangian tells us that the transition reads also \cite{CasalbuoniPG}
\beq\nonumber
\Gamma(B^*_0 \to B^+ \pi^-)=\frac{h^2}{8\pi f^2_\pi}\frac{m_B}{m^3_{B^*_0}}\left(m^2_{B^*_0}-m^2_B\right)^2 |\vec{q}_\pi|,
\eeq
by the identification 
$$g_{B^*_0 B \pi}=\sqrt{\frac{m_B}{m_{B^*_0}}}\left(m^2_{B^*_0}-m^2_B\right)\frac{h}{f_\pi} \,,$$
that is appropriate in the heavy quark limit. In the static limit, the coupling $\widetilde{g}$ is similar to $\widehat{g}$, but for hadronic transition between positive parity states. Those transitions are energetically not allowed for the $B$ system but are useful for chiral extrapolations in Lattice QCD.
So far there is only one computation of $h$ and $\widetilde{g}$ \cite{BecirevicZZA}, using ratio of three-point correlation functions and the techniques of measuring the Fourier transform of the radial distribution to obtain the form factor $A_+(q_\pi^2)$ in the limit $q_\pi^2 \to 0$, to extract $h$:
\beq\nonumber
A_+(\delta^2 - q_\pi^2) = 4 \pi \int_0^\infty r^2 dr \frac{\sin (q_\pi r)}{q_\pi r} f_{PAS}(r),
\eeq
where $\delta = m_{B^*_0}-m_B$, $f_{PAS}(r)=\langle B|[\bar{q}\gamma_0 \gamma_5 q](r)|B^*_0\rangle$ is the radial distribution depicted in Figure~\ref{figA0} and $\vec{q}_\pi=(0,0,q_z)$, choosing $q_z=\delta$.\\ 
In Ref.~\cite{McNeileRF} the transition $B^*_0 \to B \pi$ was directly studied on the lattice, computing two-point correlation functions: the authors claimed that, close to the threshold $m_{B \pi}\sim m_{B^*_0}$, the ratio 
$$C^{(2)}_{B^*_0\,B\pi}(t)/\sqrt{C^{(2)}_{B^*_0\,B^*_0}(t)C^{(2)}_{B\pi\,B\pi}(t)} \,,$$
is related to $\Gamma(B^*_0 \to B^+ \pi^-)$. We follow here this last approach and perform the computation on a set of 
${\rm N_f} = 2$ configurations made available by the Lattice Coordinated Simulations
effort. It gives a further check that the extraction of the scalar $B$ meson decay constant on 
those ensembles, that we report in a forthcoming paper, is under control at $\sim$ 10\% of precision we hope. 
The plan of the letter is the following: in section \ref{sec2} and \ref{sec3} we 
describe the approach we have employed, in section \ref{sec4} we present our lattice set-up 
and our results are given in section \ref{sec5}, that we discuss in section \ref{sec6}.

\section{\label{sec2}Extraction of $\langle B \pi|B^*_0\rangle$}

The transition amplitude under interest is parametrised by
\beq\nonumber
\langle \pi^+(q_\pi)B^-(p)|B^{*0}_0(p')\rangle=g_{B^*_0 B \pi}=\sqrt{m_B m_{B^*_0}}\frac{m^2_{B^*_0}-m^2_B}{m^2_{B^*_0}}
\frac{h}{f_\pi} \,,
\eeq
with $q_\pi=p'-p$ and $f_{\pi}=130$~MeV, the pion decay constant. When the transition amplitude is small, the Fermi golden rule teaches us that
\beq\nonumber
\Gamma(B^*_0 \to B^- \pi^+) = 2 \pi\, |\langle \pi^+(q_{\pi})B^-(p)|B^{*0}_0(p')\rangle|^2\, \rho \,,
\eeq
where the density of states $\rho$ reads, for a given energy $E_\pi$ of the pion living on the lattice of spatial volume $L^3$,
\bea\nonumber
\rho(E_{\pi}) = \frac{L^3}{(2\pi)^3}\, 4\pi \vec{q}^{\, 2}_\pi\, \frac{dq_\pi}{dE_\pi} = \frac{L^3}{2\pi^2} |\vec{q}_\pi| E_\pi \,.
\eea
In lattice units ($a$ being the lattice spacing), we obtain
\beq\nonumber
\frac{\Gamma(B^*_0 \to B^- \pi^+)}{q_\pi}=\frac{1}{\pi}\left(\frac{L}{a}\right)^3 (aE_\pi) \,
|a\langle \pi^+(q_\pi)B^-(p)|B^{*0}_0(p')\rangle|^2\,.
\eeq
Considering the two-point correlation function $C^{(2)}_{B^*_0\,B\pi}(t)=\langle \mathcal{O}^{B\pi}(t) \mathcal{O}^{B^*_0 \dag}(0) \rangle$, 
where $O^{B^*_0}$ and $O^{B\pi}$ are interpolating fields with vanishing momentum of the $B^*_0$ and the $B\pi$ states respectively, we have
\begin{align*}
C^{(2)}_{B^*_0\,B\pi}(t) &= \sum_{t_1} \langle 0 | \mathcal{O}^{B^*_0} | B^*_0 \rangle x \langle B\pi | \mathcal{O}^{B\pi} | 0 \rangle e^{-m_{\Bscal}t_1} e^{-m_{B\pi}(t-t_1)} + \mathcal{O}(x^3) + \text{excited\ states},
\end{align*}
with $x=|a\langle \pi^+(q_\pi)B^-(p)|B^{*0}_0(p')\rangle|$.
We have assumed small overlaps $ \langle 0 | \mathcal{O}^{\Bscal} | B \pi \rangle$ and  
$\langle 0 | \mathcal{O}^{B\pi} | \Bscal \rangle$ and the normalization of states is $\langle n | m\rangle = \delta_{mn}$.
Finally, close to the threshold $m_{\Bscal} \approx m_{B\pi}$, we get
\begin{align*}
C^{(2)}_{\Bscal\,B\pi}(t) &=  \langle 0 | \mathcal{O}^{\Bscal} | \Bscal \rangle x \langle B\pi | \mathcal{O}^{B\pi} | 0 \rangle \times t e^{-m_{\Bscal}t} + \mathcal{O}(x^3) + \text{excited\ states}.
\end{align*}
Therefore, one can extract $x$ from the ratio \cite{McNeileXX} - \cite{McNeileAZ}
\begin{align}
R(t) = \frac{C^{(2)}_{\Bscal\,B\pi}(t)}{\left( C^{(2)}_{\Bscal\,\Bscal}(t) C^{(2)}_{B\pi\,B\pi}(t) \right)^{1/2}} \approx A  +  xt \,,
\label{alpha}
\end{align}
where $C^{(2)}_{\Bscal\,\Bscal}$ and $C^{(2)}_{B\pi\,B\pi}$ are, respectively, two-point correlation functions of a scalar $B$ meson and a $B \pi$ multihadronic state:
\beq\nonumber
C^{(2)}_{\Bscal\,\Bscal}(t)=\langle \mathcal{O}^{\Bscal}(t) \mathcal{O}^{\Bscal \dag}(0) \rangle, \quad
C^{(2)}_{B\pi\, B\pi}(t)=\langle \mathcal{O}^{B\pi}(t) \mathcal{O}^{B\pi \dag}(0) \rangle \,.
\eeq
Further away from the threshold, eq.(1) has to be modified. The most interesting correction
for our analysis is the one to the linear term in $x$. The time dependence of the ratio $R$ is then in
\begin{equation}
t \longrightarrow \frac{ 2 }{ \Delta} \mathrm{sinh}\left( \frac{\Delta}{2}t \right) = t + \frac{\Delta^2 t^3}{24} + \mathcal{O}(\Delta^4) \,,
\label{time_dep} 
\end{equation}
where $\Delta = m_{\Bscal} - m_{B\pi}$.
\noindent To suppress the contamination by excited states, it is welcome to solve a Generalized Eigenvalue Problem (GEVP) \cite{MichaelNE} - \cite{BlossierQMA}:
\begin{align}
R^{\GEVP}(t) = \frac{\left( v_{\Bscal}(t,t_0), C^{(2)}_{\Bscal\,B\pi}(t) 
v_{B\pi}(t,t_0) \right) }{\sqrt{  \left( v_{\Bscal}(t,t_0), C^{(2)}_{\Bscal\,\Bscal}(t) v_{\Bscal}(t,t_0) \right) \times \left( v_{B\pi}(t,t_0) , C^{(2)}_{B\pi\,B\pi}(t) v_{B\pi}(t,t_0) \right) }} \,,
\label{alpha_GEVP}
\end{align}
where $C^{(2)}_{\Bscal\,B\pi}$, $C^{(2)}_{\Bscal\,\Bscal}$ and $C^{(2)}_{B\pi\,B\pi}$ are from now matrices of two-point correlators and $v_X$ are the generalized eigenvectors associated to the ground state in the corresponding channel
\begin{align}
C^{(2)}_{\Bscal\,\Bscal}(t) v_{\Bscal}(t,t_0) &= \lambda_{\Bscal}(t,t_0) C^{(2)}_{\Bscal\,\Bscal}(t_0) v_{\Bscal}(t,t_0) \,, \label{eigenvec1}\\
C^{(2)}_{B\pi\,B\pi}(t) v_{B\pi}(t,t_0) &= \lambda_{B\pi}(t,t_0) C^{(2)}_{B\pi\,B\pi}(t_0) v_{B\pi}(t,t_0) \,,
\end{align}
and $(a, C b)=\sum_i a_i C_{ij} b_j$ is the scalar product. 

\section{\label{sec3}Extraction of $\widetilde{g}$}

Similarly to the coupling $\widehat{g}$ which sets the magnitude of the transition between the pseudoscalar and the vector $B$ mesons by exchanging a single soft pion \cite{CasalbuoniPG}, the coupling $\widetilde{g}$ parametrizes the amplitude
\begin{equation}
\langle B_0^{*} \, |\, \overline{\psi}_l \gamma_k \gamma_5 \psi_l\, |\,  B_1^{*}(\epsilon_k)  \rangle = \widetilde{g} \, \epsilon_k \,,
\end{equation}
where $B_0^{*}$ and $B_1^{*}$ are respectively the scalar and the axial $B$ mesons at rest and $\epsilon_k$ is the polarization vector of the axial $B$ meson. This matrix element can be extracted using the same technique as discussed in \cite{BulavaYZ} but applied to the first excited heavy-light mesons doublet. Therefore following the method of \cite{BulavaYZ, BlossierQMA}, we consider the ratio of three to two-point correlation functions
\begin{align}
\mathcal{M}^{\rm sGEVP}(t,t_0) = -\partial_t \left(  \frac{\left(v_{B_0^*}(t,t_0), \left[  K(t)/\lambda_{B_0^*}(t,t_0) - K(t_0) \right] v_{B_1^*}(t,t_0) \right)}{  \left(v_{B_0^*}(t,t_0),C^{(2)}_{B_0^*\,B_0^*}(t_0)v_{B_0^*}(t,t_0)\right)^{1/2}  \left(v_{B_1^*}(t,t_0),C^{(2)}_{B_1^*\, B_1^*}(t_0)v_{B_1^*}(t,t_0)\right)^{1/2} } \right) \,,
\label{sGEVP_estimator}
\end{align}
where $K_{ij}(t)$ is the summed three-point correlation function
\begin{align}
K_{ij}(t) &= \sum_{t_1} C_{ij}^{(3)}(t,t_1) \,, \\
C_{ij}^{(3)}(t,t_1) &= \frac{1}{V^3} \sum_{\vec{x},\vec{y},\vec{z}} \sum_{t_x} \langle \mathcal{O}^{B_0^*}_{i}(\vec{z},t+t_x) \mathcal{A}_k(\vec{y},t_1+t_x) \mathcal{O}^{B_1^* \dag}_j(\vec{x},t_x)  \rangle \,, 
\label{three_point}
\end{align}
and $\mathcal{A}_k = Z_A \overline{\psi}_l(x) \Gamma^A_k \psi_l(x)$ is the renormalized axial current. The renormalisation constant $Z_{A}$ was determined non-perturbatively by the ALPHA Collaboration \cite{Fritzsch:2012wq}. Here, $C^{(3)}(t,t_1)$ is again a matrix of correlators and the eigenvectors $v_{B_1^*}(t,t_0)$ are defined similarly to eq.~(\ref{eigenvec1}). 
Thanks to heavy quark symmetry, the two-point correlation functions $C^{(2)}_{B_0^*\,B_0^*}$ and $C^{(2)}_{B_1^*\,B_1^*}$ are proportional and only one GEVP needs to be solved. Finally, one can show that, in the static limit of HQET \cite{BulavaYZ},
\begin{align*}
	\mathcal{M}^{\mathrm{sGEVP}}(t,t_0) \xrightarrow[t_0=t-1]{t\gg 1}  \ \widetilde{g} \ +\ \mathcal{O}\left( t e^{-\Delta_{N+1,n}t} \right) \,,
\end{align*} 
where $\Delta_{mn} = E_m - E_n$ is the energy difference between the $m^{\rm th}$ and $n^{\rm th}$ excited states of the GEVP and $N \times N$ is the size of the matrix of correlators defining the GEVP. 

\section{\label{sec4}Lattice setup}

\begin{figure}[t]
	\begin{minipage}[c]{0.49\linewidth}
	\centering 
	\includegraphics*[width=\linewidth]{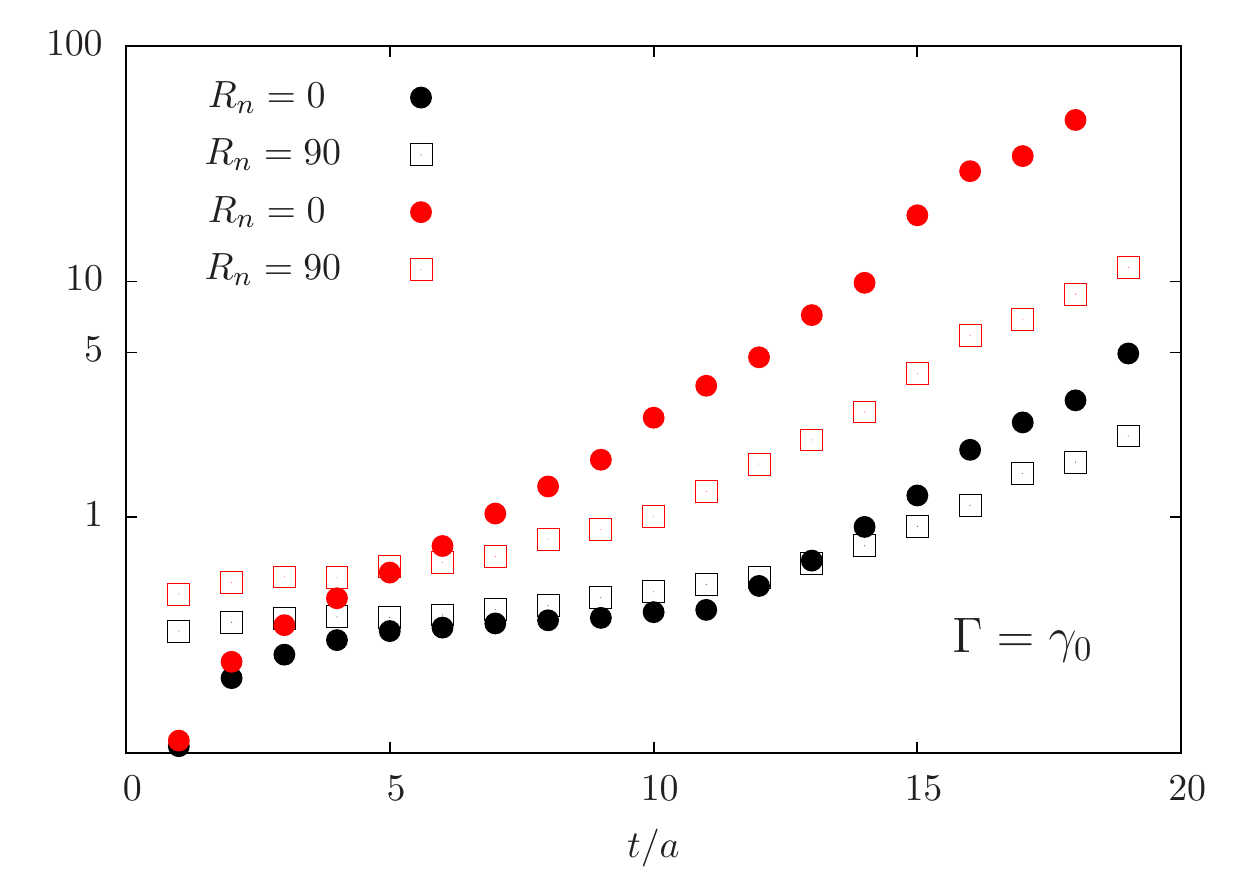}
	\end{minipage}
  	 \begin{minipage}[c]{0.49\linewidth}   
	\centering 
	\includegraphics*[width=\linewidth]{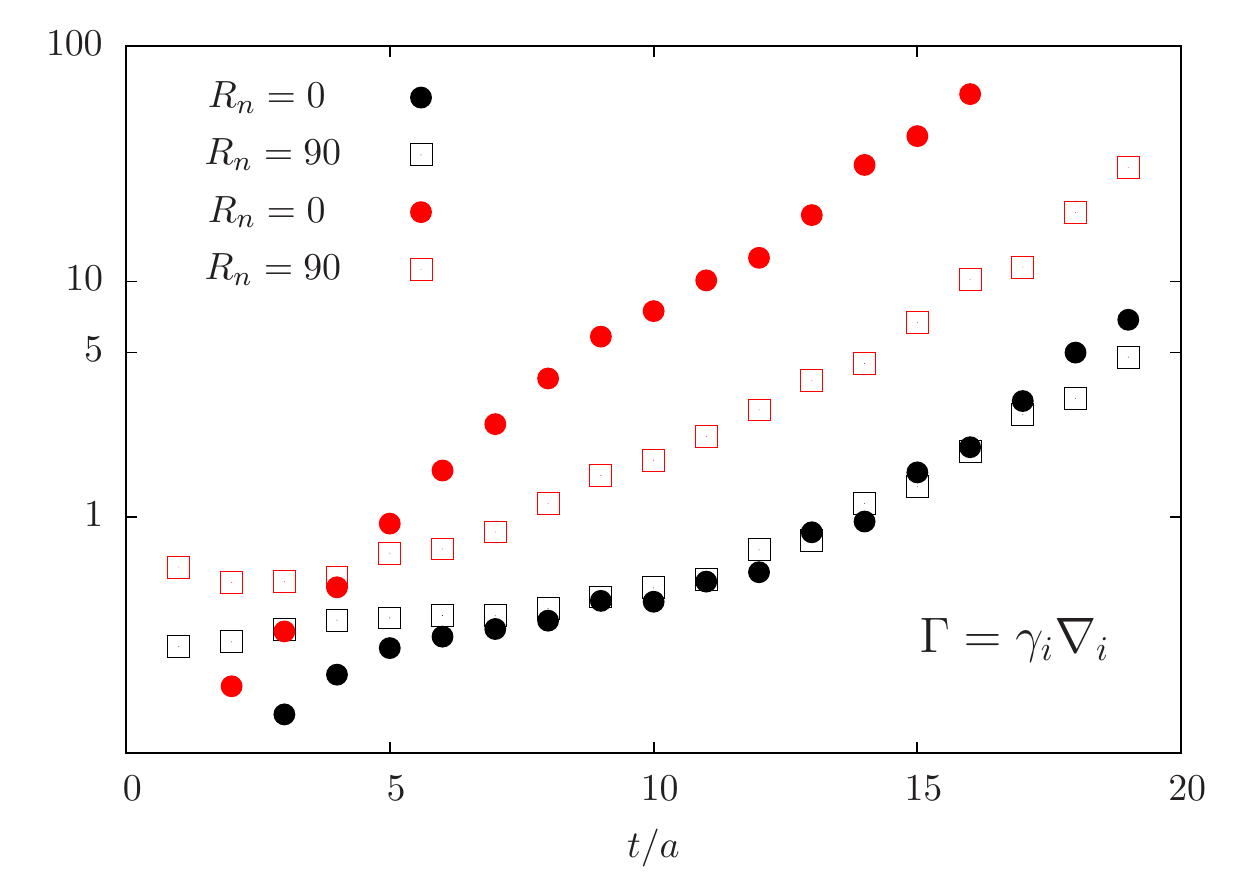}
	\end{minipage}
	\caption{Statistical error (in percent) for the correlation functions $C_{\Bscal\,B\pi}(t)$ for the two different methods explained in the text. The black points correspond to the correlators $C_{B^*_0\,B\pi}(t)$ computed using the one-end-trick and the red points correspond to the correlators $C_{B^*_0\,B\pi}(t)$ computed by inverting twice the Dirac operator. On the left, for $\Gamma=\gamma_0$ and on the right, for $\Gamma = \gamma_i \nabla_i$. The results correspond to the lattice ensemble E5.}
	\label{fig_err_cor}
\end{figure}

In our study we have performed measurements on a subset of four ${\rm N_f}=2$ CLS lattice  simulations, which have been generated using either the DD-HMC algorithm \cite{LuscherQA} - \cite{Luscherweb} or the MP-HMC algorithm \cite{MarinkovicEG}, defined with the plaquette gauge action and non perturbatively ${\cal O}(a)$ improved Wilson-Clover fermions; we collect the main parameters in Table \ref{tabsim} and we remind the reader that the criterion of our choice is to be very close to the threshold $m_{\Bscal} \approx m_{B\pi}$ (see Table \ref{tab:resultsh}). We have computed static-light correlators with HYP2 static quarks \cite{HasenfratzHP} and stochastic all-to-all propagators with full time dilution for the light quarks \cite{FoleyAC}. A single stochastic source has been used to compute the propagator. Interpolating fields of a static-light meson are defined as \cite{GuskenAD}
 \renewcommand{\arraystretch}{1.5}
 \setlength{\tabcolsep}{0.2cm}
\begin{table}[t]
\begin{center}
\begin{tabular}{|c|c|c|c|c|c|c|c|}
\hline
CLS label	&	$\beta$	&	$L^3\times T$	&	$\kappa$	&	$a$[fm]	&	$m_\pi$[MeV]	&	$\#_h$  &  $\#_{\widetilde{g}}$ \\
\hline
B6		&	5.2		&	$48^3\times 96$&0.13597&0.075&280&250 & 200\\
\hline
E5		&	5.3		&	$32^3\times 64$&0.13625&0.065&440&450 & 400\\
F6		&			&	$48^3 \times 96$&0.13635&&310&300 & 250\\
\hline
N6		&	5.5		&	$48^3\times 96$&0.13667&0.048&340&250 & 200\\
\hline
\end{tabular}
\end{center}
\caption{\label{tabsim} Simulations parameters: the bare coupling constant $\beta = 6/g_0^2$, spatial extent in lattice units $L$ (with $T = 2L$), hopping parameter $\kappa$, lattice spacing $a$ in physical units, pion mass $m_{\pi}$ and number of configurations used for the computation of the two-point and three-point correlation functions respectively. }
\end{table} 
\begin{equation}
O^B_{\Gamma,n} = \overline{\psi}_l^{(n)} \Gamma \psi_h \quad , \quad \psi_l^{(n)} \equiv (1 + \kappa_G a^2 \Delta)^{R_n} \psi_l \,,
\end{equation}
where $\psi_h$ is the static heavy quark field and $\psi_l$ is the relativistic quark field ($l=u/d$). The Gaussian smearing parameters are $\kappa_G=0.1$, $r_n\equiv 2a \sqrt{\kappa_G R_n} \leq 0.6$~fm and $\Delta$ is a covariant Laplancian made of three times APE-blocked links \cite{AlbaneseDS}. Moreover, $O_{\Gamma,n}$ can be ``local'' ($\Gamma=\gamma_0, \gamma_5$) or contain a derivative operator ($\Gamma=\gamma_0 \sum_{i=1}^3 \gamma_i \nabla_i$, $\Gamma=\gamma_5 \sum_{i=1}^3 \gamma_i \nabla_i$) where $\nabla_i$ is the symmetrized covariant derivative acting on the light quark field: 
$\nabla_i \psi_l(x) = ( U_i(x) \psi_l(x+\hat{\i})-U^\dag_i(x) \psi_i(x-\hat{\i}) ) /2$.
We have also implemented the isosymmetric interpolating fields of the form 
\begin{multline*}
\mathcal{O}^{B\pi}_{n} = \frac{1}{V^2} \sum_{\vec{x}_1,\vec{x}_2} \sqrt{\frac{2}{3}} \left[ \overline{\psi}_d(x_1) \gamma_5 \psi_u(x_1) \right]  \left[ \overline{\psi}_u^{(n)}(x_2) \gamma_5 \psi_h(x_2) \right]  \\ 
- \sqrt{\frac{1}{6}}  \left[ \overline{\psi}_u(x_1) \gamma_5 \psi_u(x_1) - \overline{\psi}_d(x_1) \gamma_5 \psi_d(x_1) \right] \left[ \overline{\psi}_d^{(n)}(x_2) \gamma_5 \psi_h(x_2) \right] \,, 
\end{multline*}
which couple to the multihadronic state 
$$ \sqrt{\frac{2}{3}} \pi^+(0) B^{-}(0) - \sqrt{\frac{1}{3}} \pi^0(0) \overline{B}^0(0) \,.$$
Using the notation 
$\contraction[1ex]{}{\psi}{_l^{(m)}(x)}{\overline{\psi}}   \psi_l^{(m)}(x)\overline{\psi}_l^{(n)}(y) = G^{mn}_l(x,y)$, 
$\contraction[1ex]{}{\psi_h}{(x)}{\overline{\psi}} \psi_h(x)\overline{\psi}_h(y) = G_h(x,y)$ 
for the smeared light quark propagator and the static quark propagator respectively, the two-point correlation functions constructed from these interpolating fields are
\begin{align}
C_{B^*_0\,B^*_0}^{nm}(t) &= - \frac{1}{V^2}  \sum_{\vec{x},\vec{y}} \ \mathrm{Tr}\left[  G^{mn}_l(y,x)  \Gamma_1  G_h(x,y) \overline{\Gamma}_2 \right] 
\label{CBB},
\end{align}
$\overline{\Gamma}=\gamma_0 \Gamma^\dag \gamma_0$, whose the diagram is sketched in Figure~\ref{figdiagB},
\begin{align}
C_{B\pi\,B\pi}^{nm}(t) &= \frac{1}{V^4}  \sum_{\vec{x}_i,\vec{y}_i} \ \mathrm{Tr}\left[  G_l(y_1,x_1)  \gamma_5  G_l(x_1,y_1) \gamma_5  \right] \times  \mathrm{Tr}\left[  G_h(y_2,x_2)  \gamma_5  G_l^{nm}(x_2,y_2) \gamma_5 \right]  \label{BpiBpi1}\\
&\quad - \frac{3}{2V^4}  \sum_{\vec{x}_i,\vec{y}_i} \ \mathrm{Tr}\left[  G_l(y_1,x_1)  \gamma_5  G_l^{0n}(x_1,x_2) \gamma_5  G_h(x_2,y_2)  \gamma_5  G_l^{m0}(y_2,y_1) \gamma_5 \right] \label{BpiBpi2}\\
&\quad + \frac{1}{2V^4}  \sum_{\vec{x}_i,\vec{y}_i} \ \mathrm{Tr}\left[  G_l^{0n}(y_1,x_2)  \gamma_5  G_h(x_2,y_2) \gamma_5  G_l^{m0}(y_2,x_1)  \gamma_5  G_l(x_1,y_1) \gamma_5 \right]  \,, \label{BpiBpi3}
\end{align}
whose the direct (\ref{BpiBpi1}), box (\ref{BpiBpi2}) and cross (\ref{BpiBpi3}) diagrams are sketched in Figure~\ref{figdiagBpiBpi},
\begin{figure}[t]

	\centering
	\vspace{-1cm}
	\unitlength = 0.8mm
	\begin{fmffile}{feyn/B_B}
	\begin{fmfgraph*}(60,40)
	\fmfleft{i}
	\fmfright{o}
	\fmf{phantom,tension=5}{i,v1}
	\fmf{phantom,tension=5}{v2,o}
	\fmf{dbl_plain_arrow}{v1,v2}
	\fmf{fermion,left,tension=0.4}{v2,v1}
	\fmfv{decor.shape=circle,decor.filled=full, decor.size=4thick}{v1}
	\fmfv{decor.shape=circle,decor.filled=full, decor.size=4thick}{v2}
	\fmflabel{$y, \Gamma_2$}{v1}
	\fmflabel{$x,\Gamma_1$}{v2}
	\end{fmfgraph*}
	\end{fmffile}
	\caption{Diagram representing the correlator $C_{B^*_0\,B^*_0}^{nm}$. The simple and double lines represent the light and static quark propagators respectively.}
	\label{figdiagB}
	
	\vspace{1cm}

	\begin{minipage}[c]{0.32\linewidth}

	\vspace{-1cm}
	\unitlength = 0.8mm
	\begin{fmffile}{feyn/Bpi_Bpi_direct}
	\begin{fmfgraph*}(60,20)
	\fmfleft{i1,i2}
	\fmfright{o1,o2}
	\fmf{phantom,tension=5}{i1,v1}
	\fmf{phantom,tension=5}{v2,o1}
	\fmf{phantom,tension=5}{i2,v3}
	\fmf{phantom,tension=5}{v4,o2}
	\fmf{dbl_plain_arrow}{v1,v2}
	\fmf{fermion,left,tension=0.3}{v2,v1}
	\fmf{fermion}{v3,v4}
	\fmf{fermion,left,tension=0.3}{v4,v3}
	\fmfv{decor.shape=circle,decor.filled=full, decor.size=4thick}{v1}
	\fmfv{decor.shape=circle,decor.filled=full, decor.size=4thick}{v2}
	\fmfv{decor.shape=circle,decor.filled=full, decor.size=4thick}{v3}
	\fmfv{decor.shape=circle,decor.filled=full, decor.size=4thick}{v4}
	\fmflabel{$y_1$}{v3}
	\fmflabel{$y_2$}{v1}
	\fmflabel{$x_1$}{v4}
	\fmflabel{$x_2$}{v2}
	\end{fmfgraph*}
	\end{fmffile}	
	
	\end{minipage}
	\begin{minipage}[c]{0.32\linewidth}
		
	\unitlength = 0.8mm
	\begin{fmffile}{feyn/Bpi_Bpi_box}
	\begin{fmfgraph*}(60,40)
	\fmfleft{i1,i2}
	\fmfright{o1,o2}
	\fmf{phantom,tension=5}{i1,v1}
	\fmf{phantom,tension=5}{v2,o1}
	\fmf{phantom,tension=5}{v3,o2}
	\fmf{phantom,tension=5}{i2,v4}
	\fmf{dbl_plain_arrow}{v1,v2}
	\fmf{fermion}{v2,v3}
	\fmf{fermion}{v3,v4}
	\fmf{fermion}{v4,v1}
	\fmfv{decor.shape=circle,decor.filled=full, decor.size=4thick}{v1}
	\fmfv{decor.shape=circle,decor.filled=full, decor.size=4thick}{v2}
	\fmfv{decor.shape=circle,decor.filled=full, decor.size=4thick}{v3}
	\fmfv{decor.shape=circle,decor.filled=full, decor.size=4thick}{v4}
	\fmflabel{$y_1$}{v4}
	\fmflabel{$y_2$}{v1}
	\fmflabel{$x_1$}{v3}
	\fmflabel{$x_2$}{v2}
	\end{fmfgraph*}
	\end{fmffile}	
	
	\end{minipage}
	\begin{minipage}[c]{0.32\linewidth}
	
	\unitlength = 0.8mm
	\begin{fmffile}{feyn/Bpi_Bpi_sablier}
	\begin{fmfgraph*}(60,40)
	\fmfleft{i1,i2}
	\fmfright{o1,o2}
	\fmf{phantom,tension=5}{i1,v1}
	\fmf{phantom,tension=5}{v2,o1}
	\fmf{phantom,tension=5}{v3,o2}
	\fmf{phantom,tension=5}{i2,v4}
	\fmf{dbl_plain_arrow}{v1,v2}
	\fmf{fermion}{v2,v4}
	\fmf{fermion}{v4,v3}
	\fmf{fermion}{v3,v1}
	\fmfv{decor.shape=circle,decor.filled=full, decor.size=4thick}{v1}
	\fmfv{decor.shape=circle,decor.filled=full, decor.size=4thick}{v2}
	\fmfv{decor.shape=circle,decor.filled=full, decor.size=4thick}{v3}
	\fmfv{decor.shape=circle,decor.filled=full, decor.size=4thick}{v4}
	\fmflabel{$y_1$}{v4}
	\fmflabel{$y_2$}{v1}
	\fmflabel{$x_1$}{v3}
	\fmflabel{$x_2$}{v2}
	\end{fmfgraph*}
	\end{fmffile}	

	\vspace{0.3cm}
	\end{minipage}

	\caption{\label{figdiagBpiBpi} Direct, box and cross diagrams contributing to the correlator $C_{B\pi\,B\pi}^{nm}$.}
	\label{Diag_quatre_quark}
	
	\vspace{1cm}
	
	\begin{minipage}[c]{0.49\linewidth}
		
	\unitlength = 0.8mm
	\begin{fmffile}{feyn/B_Bpi}
	\begin{fmfgraph*}(60,40)
	\fmfleft{i2,i1}
	\fmfright{o2,o1}
	\fmf{phantom,tension=5}{i1,v1}
	\fmf{phantom,tension=5}{i2,v4}
	\fmf{phantom,tension=5}{v2,o1}
	\fmf{phantom,tension=5}{v3,o2}
	\fmf{dbl_plain_arrow}{v1,v2}
	\fmf{fermion}{v2,v3}
	\fmf{fermion}{v3,v1}
	\fmfv{decor.shape=circle,decor.filled=full, decor.size=4thick}{v1}
	\fmfv{decor.shape=circle,decor.filled=full, decor.size=4thick}{v2}
	\fmfv{decor.shape=circle,decor.filled=full, decor.size=4thick}{v3}
	\fmflabel{$y, \Gamma$}{v1}
	\fmflabel{$x_1, \gamma_5$}{v3}
	\fmflabel{$x_2,\gamma_5$}{v2}
	\end{fmfgraph*}
	\end{fmffile}	
	
	\end{minipage}
	\begin{minipage}[c]{0.49\linewidth}

	\unitlength = 0.8mm
	\begin{fmffile}{feyn/Bpi_B}
	\begin{fmfgraph*}(60,40)
	\fmfleft{i2,i1}
	\fmfright{o2,o1}
	\fmf{phantom,tension=5}{i1,v1}
	\fmf{phantom,tension=5}{i2,v4}
	\fmf{phantom,tension=5}{v2,o1}
	\fmf{phantom,tension=5}{v3,o2}
	\fmf{dbl_plain_arrow}{v1,v2}
	\fmf{fermion}{v2,v4}
	\fmf{fermion}{v4,v1}
	\fmfv{decor.shape=circle,decor.filled=full, decor.size=4thick}{v1}
	\fmfv{decor.shape=circle,decor.filled=full, decor.size=4thick}{v2}
	\fmfv{decor.shape=circle,decor.filled=full, decor.size=4thick}{v4}
	\fmflabel{$y_1,\gamma_5$}{v4}
	\fmflabel{$y_2,\gamma_5$}{v1}
	\fmflabel{$x,\Gamma$}{v2}
	\end{fmfgraph*}
	\end{fmffile}	

	\end{minipage}

	\caption{\label{figdiagBBpi} Triangle diagrams contributing to the correŽlators $C_{B\pi\,\Bscal}^{nm}$ and 
	$C_{\Bscal\,B\pi}^{nm}$.}
\end{figure}   
\begin{align}
C_{B\pi\,\Bscal}^{nm}(t) &= - \frac{1}{V^3} \sqrt{ \frac{3}{2} }  \sum_{\vec{x}_i,\vec{y}} \mathrm{Tr} \left[ G_l^{m0}(y,x_1) \gamma_5 G_l^{0n}(x_1,x_2) \gamma_5 G_h(x_2,y) \overline{\Gamma} \right] \label{BpiB} \,, \\
C_{\Bscal\,B\pi}^{nm}(t) &= - \frac{1}{V^3} \sqrt{ \frac{3}{2} }\sum_{\vec{y}_i,\vec{x}} \mathrm{Tr} \left[ G_l^{m0}(y_2,y_1) \gamma_5 G_l^{0n}(y_1,x) \Gamma G_h(x,y_2) \gamma_5 \right] \label{BBpi} \,,
\end{align}
whose the diagrams are sketched in Figure~\ref{figdiagBBpi}.
We have computed the triangle correlators $C_{B^*_0\,B\pi}$ and  $C_{B\pi\,B^*_0}$ by two methods, either using the one-end-trick and a single inversion to obtain the two light propagators \cite{FosterWU,McNeileBZ}, or getting the second light propagator by solving the Dirac equation with the first light propagator taken as a generalised source. The second approach is more noisy, as shown on Figure~\ref{fig_err_cor}. 

The box (\ref{BpiBpi2}) and cross (\ref{BpiBpi3}) diagrams depicted in Figure~\ref{Diag_quatre_quark} require at least one more inversion of the Dirac operator for each time slice and are therefore expensive to compute. They have been computed only in the case of the ensemble E5. Their contributions are small compared to the direct one given by (\ref{BpiBpi1}), 0.1\% and 1\%, respectively. Neglecting them, we obtain $ax=0.0241(10)$ whereas we obtain $ax = 0.0228(10)$ when they are taken into account. The two results are compatible within our errors and the computation of these diagrams does not seem necessary at our level of precision. Since we don't expect the light quark mass dependence to play a major role on that specific point, 
we neglect these diagrams in our calculation on other ensembles.

Finally, we have also computed the three point correlation functions (\ref{three_point}) needed for the extraction of the coupling $\widetilde{g}$ using the same basis of interpolating operators: 
\begin{align*}
C^{(3)}(t,t_1) = - \frac{Z_A}{V^3} \sum_{\vec{x},\vec{y},\vec{z}} \sum_{t_x} \frac{1}{3} \sum_{k=1}^3 \mathrm{Tr}\left[  G_h(x,z) \Gamma^A_k G_l^{n0}(z,y) \gamma_{k} \gamma_5 G_l^{0m}(y,x)  \overline{\Gamma}^S \right] \,,
\end{align*}
where $\Gamma^{S} = \gamma_0,\gamma_i \overleftarrow{\nabla_i}$ and $\Gamma^{A}_i = \gamma_5 \gamma_i, \gamma_5 \overleftarrow{\nabla_i}$.

\section{\label{sec5} Results}

\begin{figure}[t]
	\centering 
	\begin{minipage}[c]{0.49\linewidth}
	\centering 
	\includegraphics*[width=0.9\linewidth]{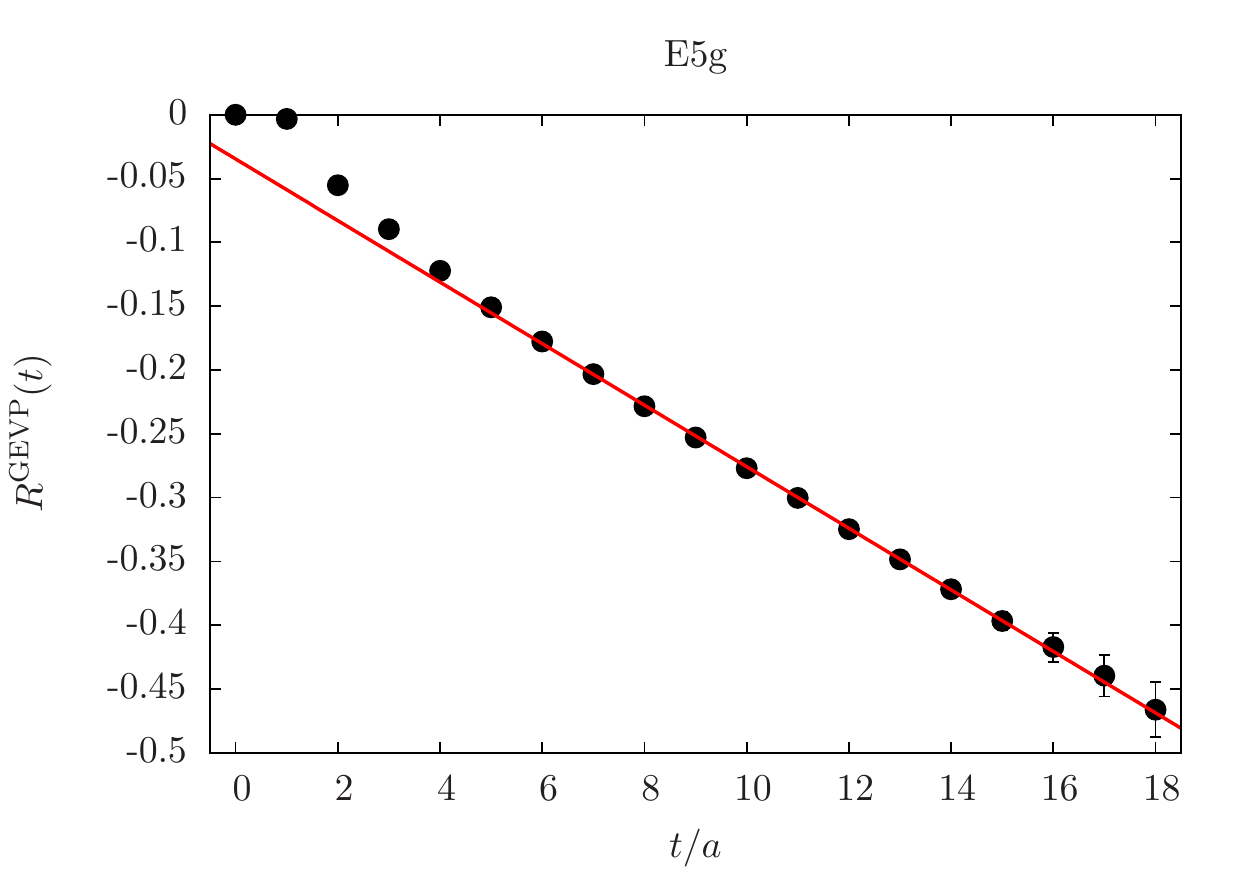}
	\end{minipage}
	\begin{minipage}[c]{0.49\linewidth}
	\centering 
	\includegraphics*[width=0.9\linewidth]{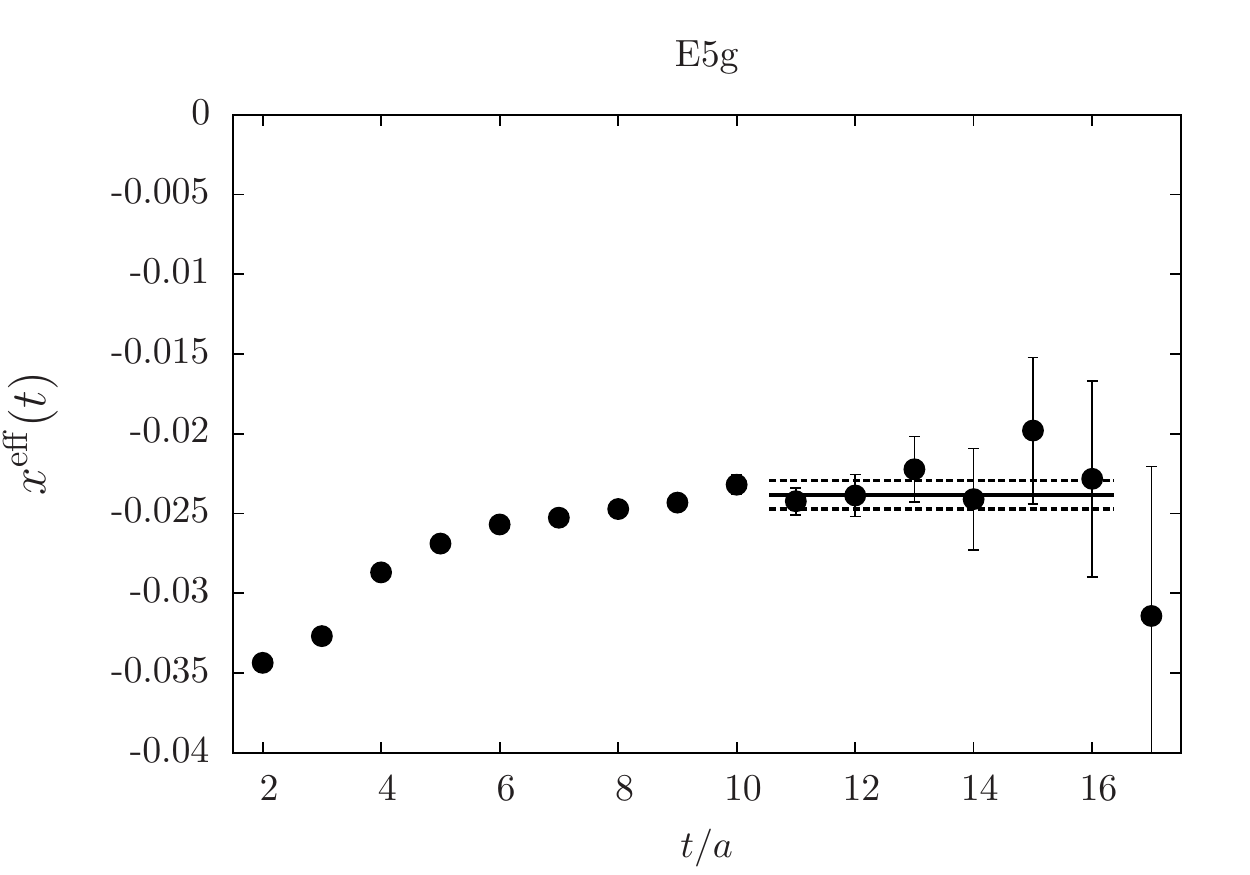}
	\end{minipage}
	\begin{minipage}[c]{0.49\linewidth}
	\centering 
	\includegraphics*[width=0.9\linewidth]{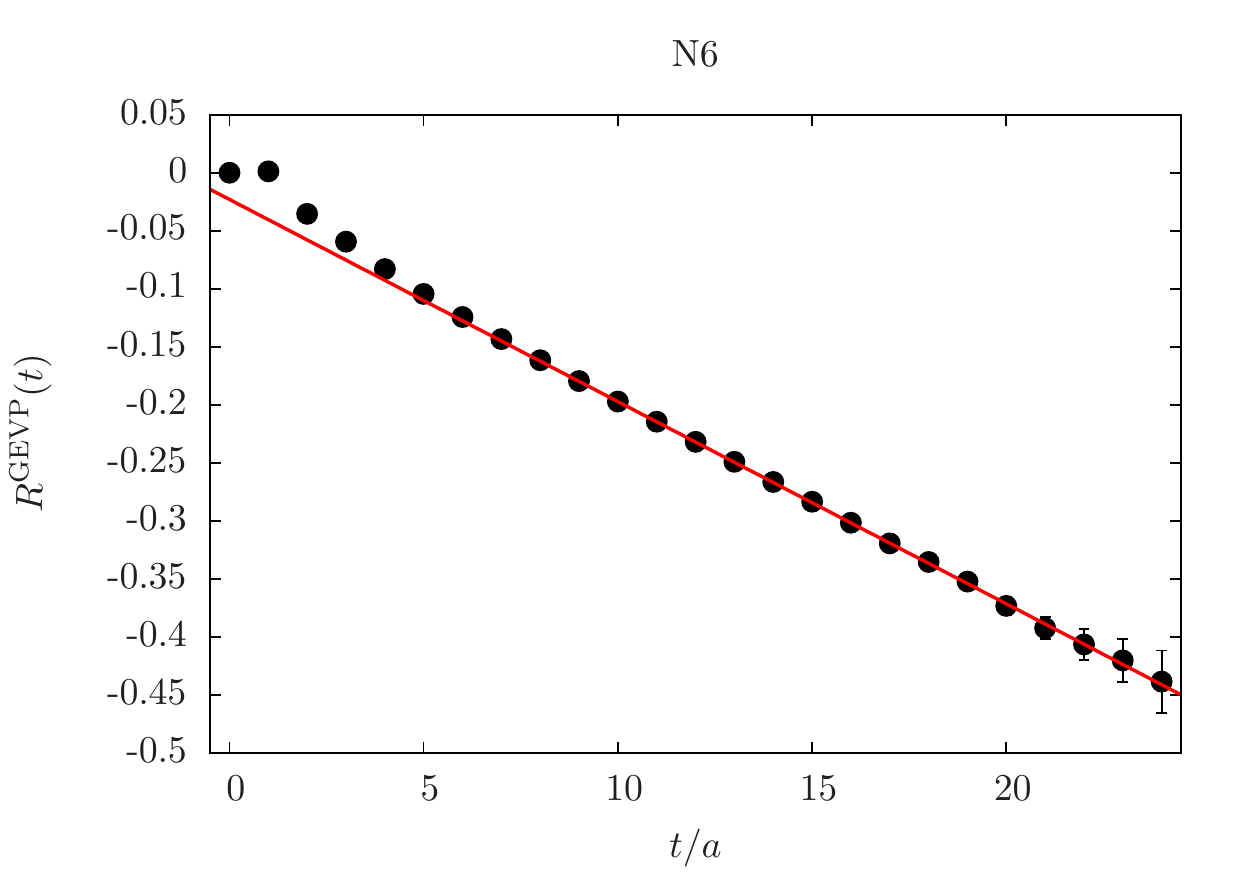}
	\end{minipage}
	\begin{minipage}[c]{0.49\linewidth}
	\centering 
	\includegraphics*[width=0.9\linewidth]{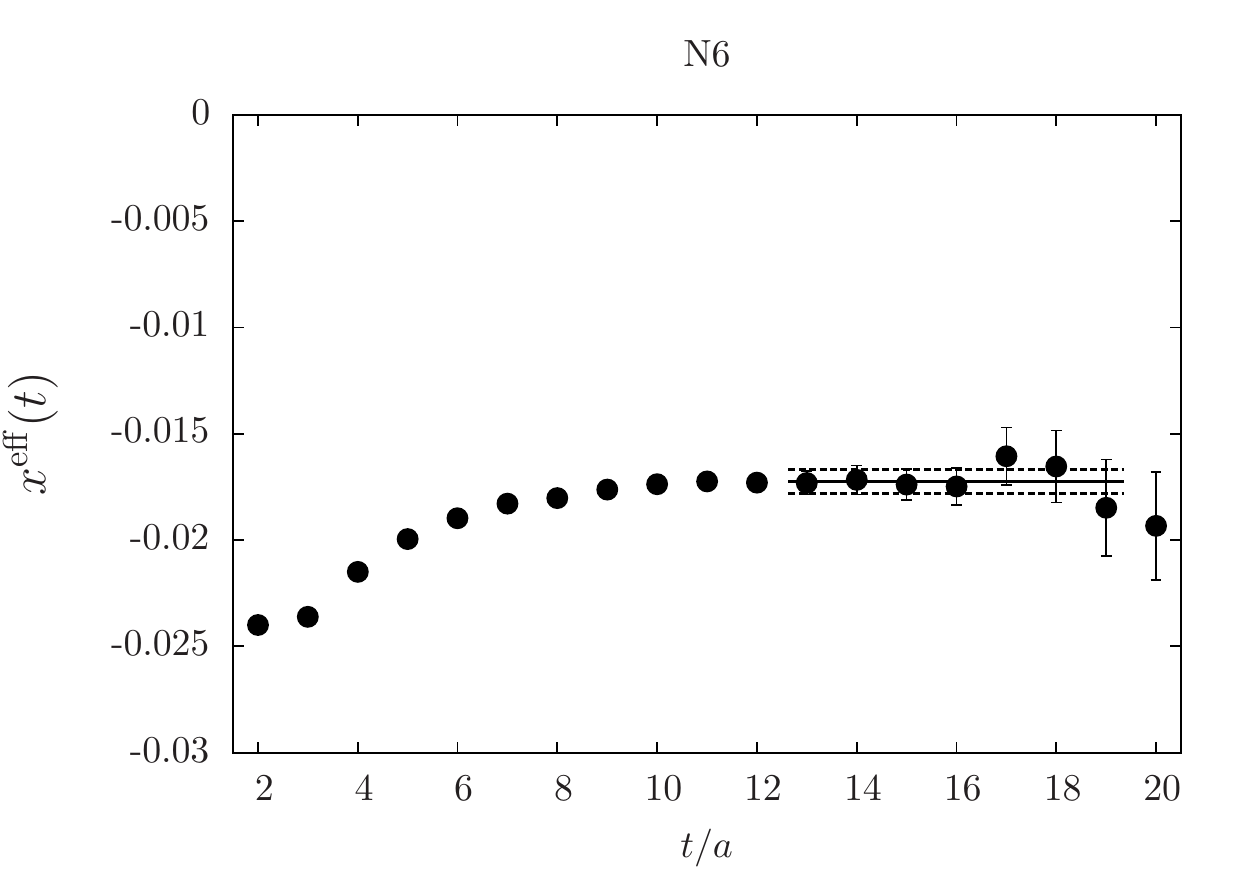}
	\end{minipage}	
	\caption{On the left: evolution of $R^{\GEVP}(t)$ with $t/a$ for the CLS ensembles E5 and N6. The red line corresponds to a linear fit where the excited states contribution is negligible. On the right: the corresponding plateaus for $x^{\rm eff}(t)$. We used $t_0/a=5$ for $t>t_0$ and $t_0=t-a$ elsewhere. }
	\label{fig:R_GEVP}
\end{figure}

%
We show in Figure~\ref{fig:R_GEVP} the ratio $R^\GEVP(t)$ and its derivative with respect to time $x^{\rm eff}(t) = dR^{\GEVP}(t)/dt$, that corresponds to the quantity $ax$ we are measuring. We observe a nice plateau for every ensemble under study. 
The very flat behavior of $x^{\rm eff}(t)$ in the plateau region lets us conclude that quadratic and higher terms in $t$ in the formula eq.~(\ref{time_dep}), coming from $\Delta \neq 0$ (see Table~\ref{tab:resultsh}), are almost absent. This was expected since in our range of fitting,
\begin{equation*}
\frac{3t^2 \Delta^2}{24} \ll 1 \quad \text{for} \quad t/a \in [0-20] \,.
\end{equation*}

\begin{figure}[t]
	\begin{minipage}[c]{0.49\linewidth}
	\centering 
	\includegraphics*[width=0.9\linewidth]{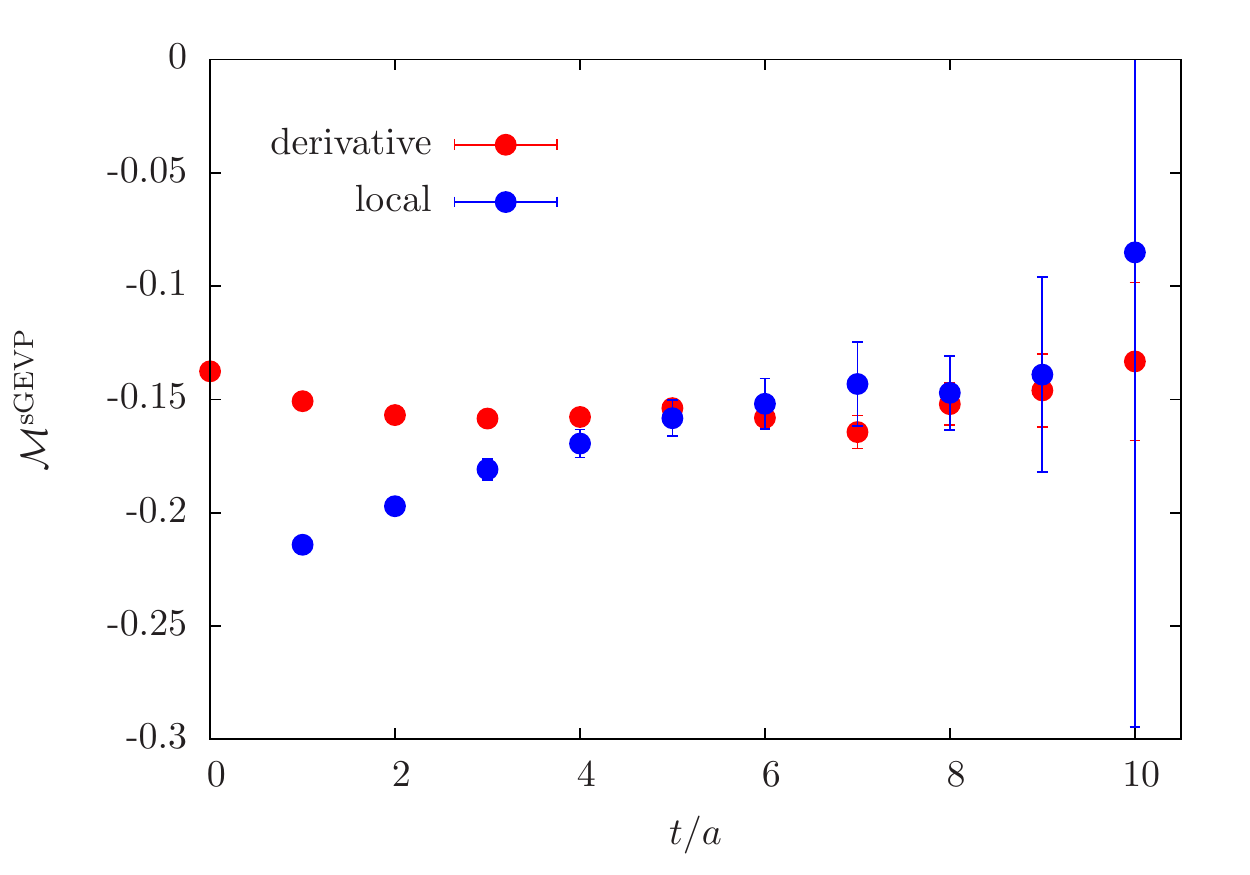}
	\end{minipage}
  	 \begin{minipage}[c]{0.49\linewidth}   
	\centering 
	\includegraphics*[width=0.9\linewidth]{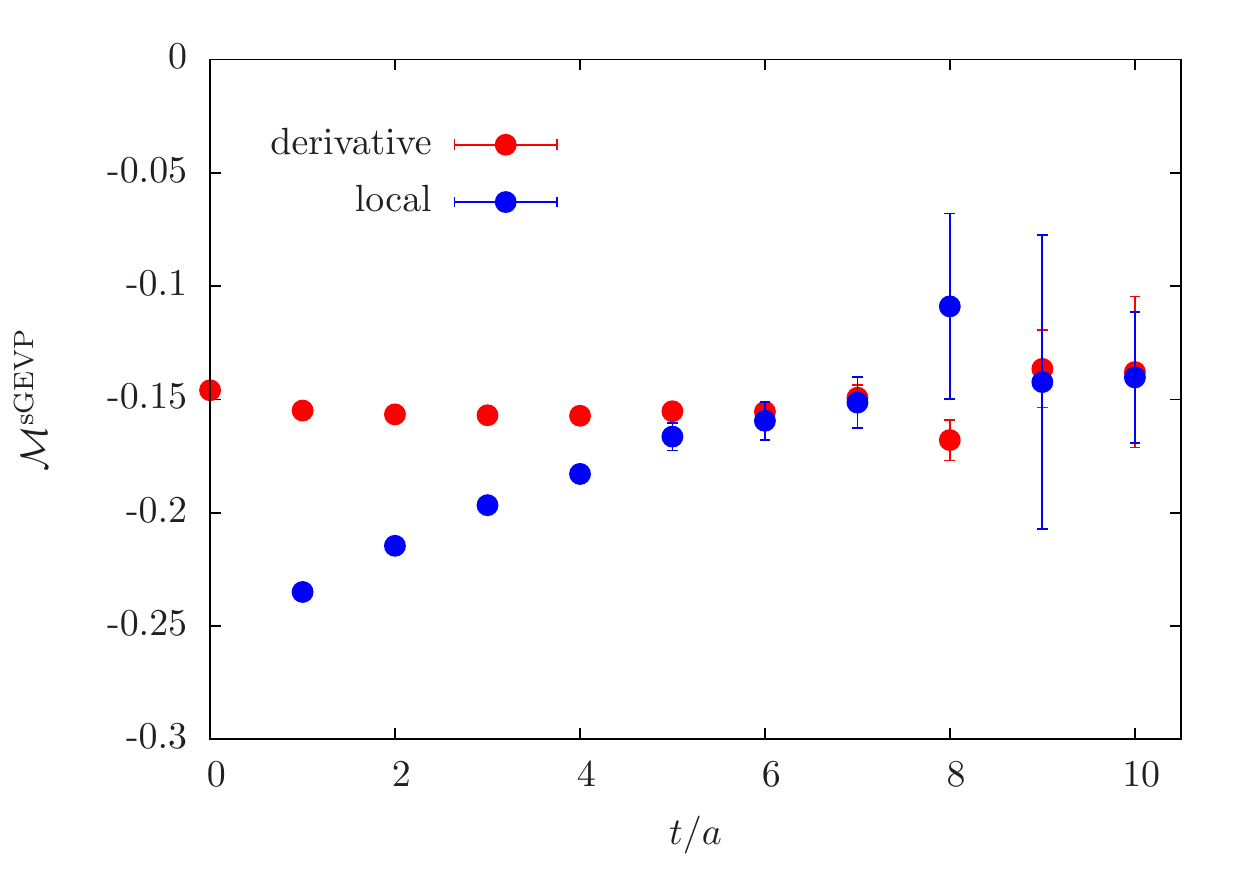}
	\end{minipage}
	\caption{Comparison of the signal obtained for $\mathcal{M}^{\text{sGEVP}}$ using local (blue) and derivative (blue) interpolating operators for the ensembles E5 (left) and F6 (right). }
	\label{g_loc_der}
\end{figure}

\begin{figure}[t]
	\centering 
	\begin{minipage}[c]{0.49\linewidth}
	\centering 
	\includegraphics*[width=0.9\linewidth]{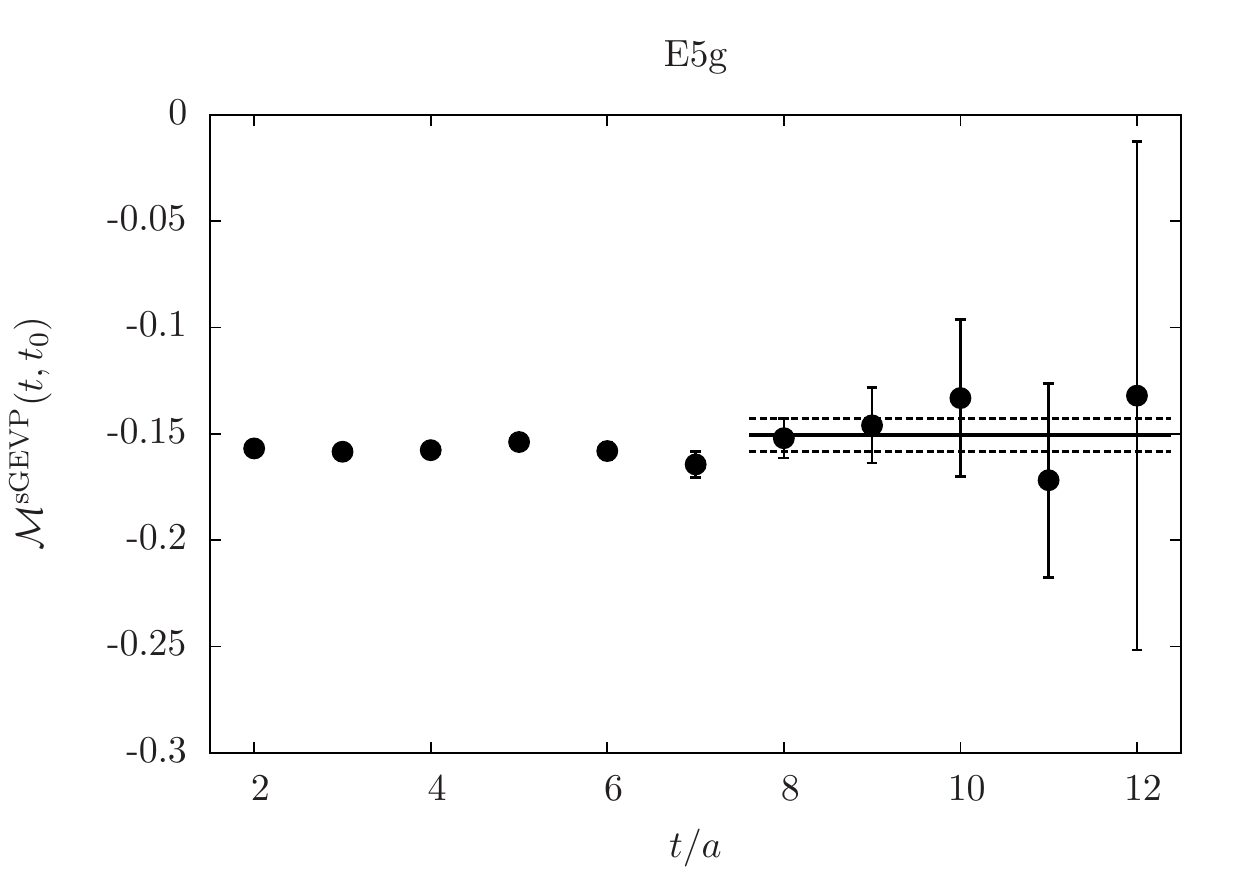}
	\end{minipage}
	\begin{minipage}[c]{0.49\linewidth}
	\centering 
	\includegraphics*[width=0.9\linewidth]{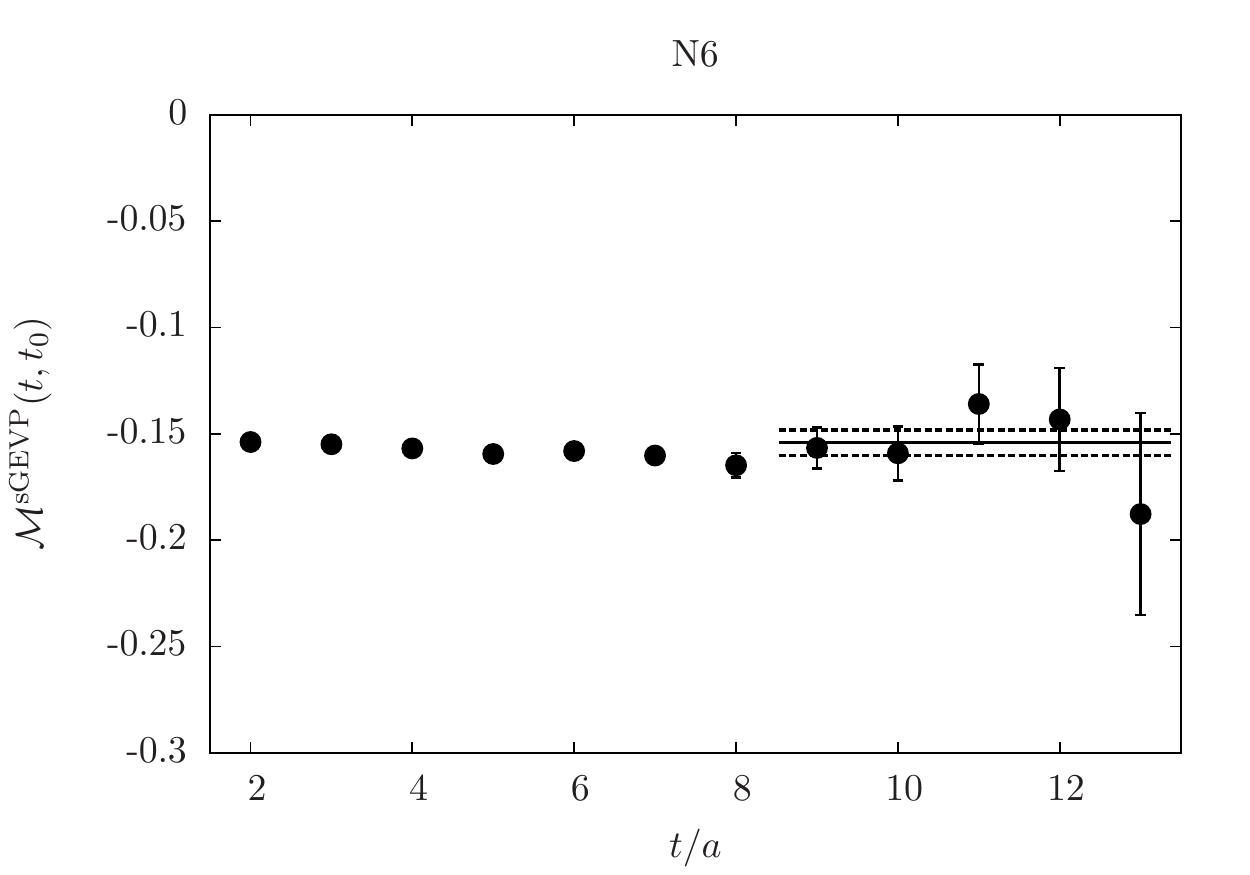}
	\end{minipage}
	\caption{Plateaus for $\mathcal{M}^{\text{sGEVP}}$ using eq.~(\ref{sGEVP_estimator}) for the CLS ensemble E5 (left) and N6 (right).}
	\label{fig:plateaugtilde}
\end{figure}

\noindent Concerning the three-point correlation functions, 
we have checked using either local interpolating operators or interpolating
operators built from the insertion of a covariant
derivative give compatible results. However in the last case the signal is less noisy as shown in Figure~\ref{g_loc_der}. Therefore only these fields are used in the following and some typical plateaus are depicted in Figure~\ref{fig:plateaugtilde}.

\begin{figure}[t]
	\centering 
	\begin{minipage}[c]{0.49\linewidth}
	\centering 
	\includegraphics*[width=0.9\linewidth]{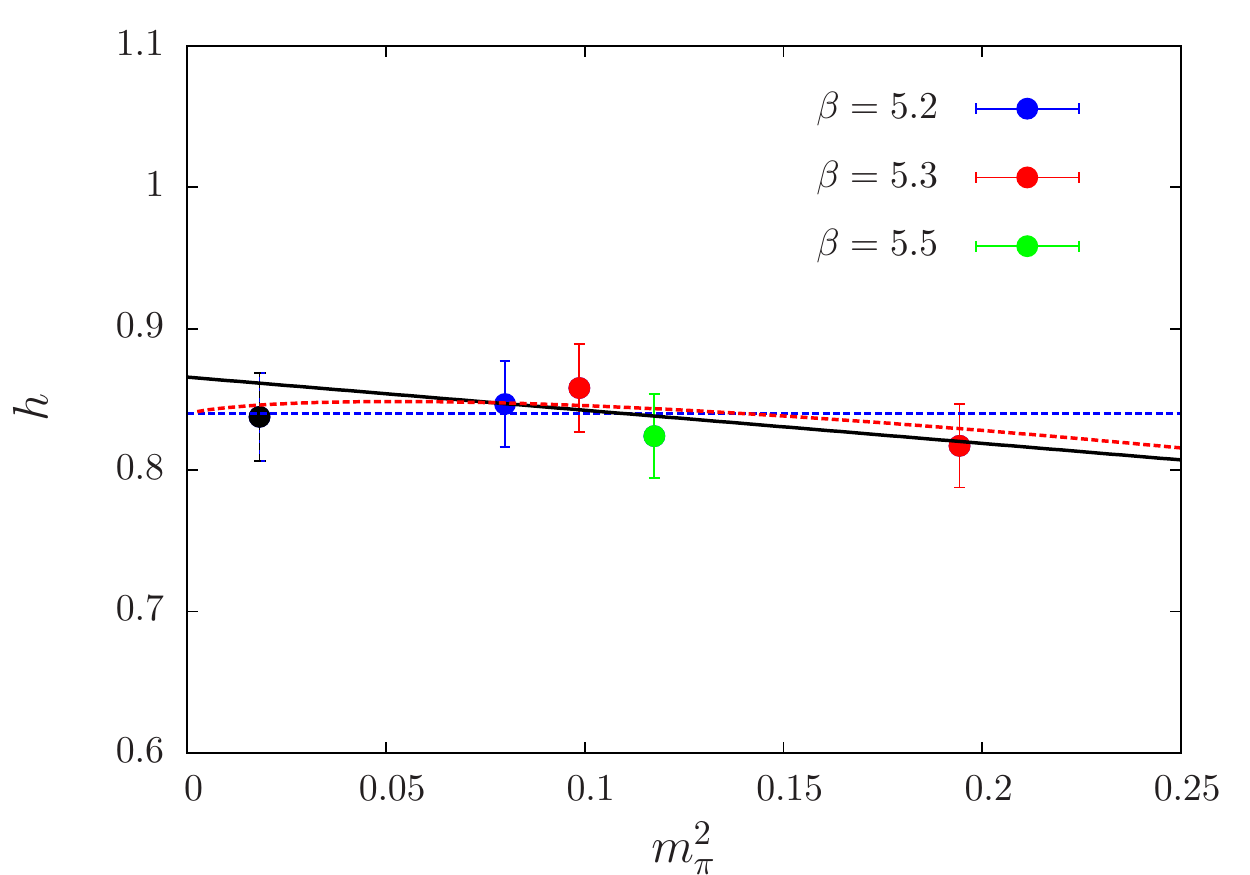}
	\end{minipage}
	\begin{minipage}[c]{0.49\linewidth}
	\centering 
	\includegraphics*[width=0.9\linewidth]{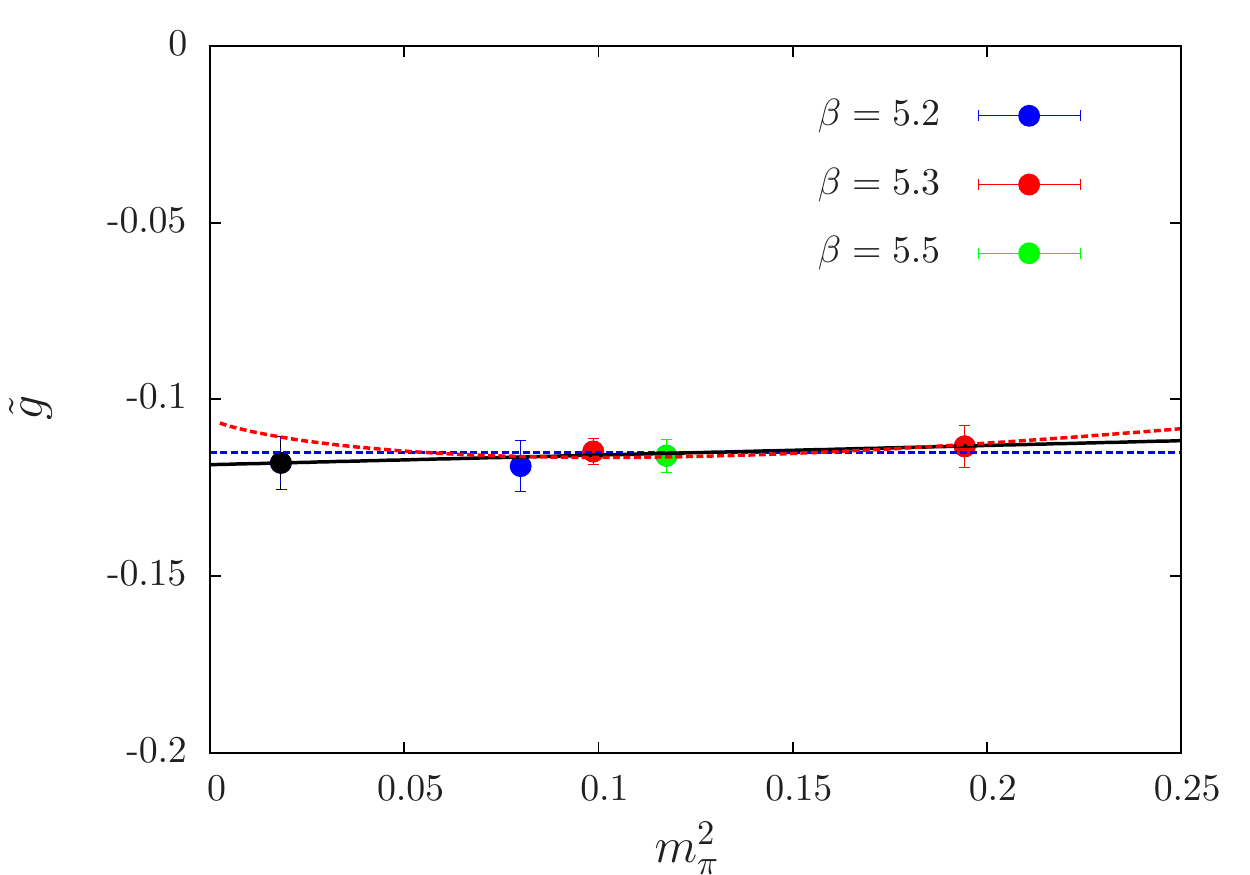}
	\end{minipage}
	\caption{Chiral extrapolations of the effective couplings $h$ (left) and $\widetilde{g}$ (right). The dashed blue line corresponds to the constant fit, the black line corresponds to the linear fit and the dashed red line corresponds to the fit formulae (\ref{fith:NLO1}), (\ref{fitg:NLO1}) with the expression derived in HM$\chi$PT.}
	\label{fig:chih}
\end{figure}

\noindent With $ax$ and $m_{\Bscal} - m_B= 385(17)_{\rm stat}(28)_{\rm syst}$~MeV \cite{spectrum2}, we can finally extract $\Gamma/|\vec{q}_\pi|$, $h$ and $g_{\Bscal B\pi}$; we collect the values in Table \ref{tab:resultsh}.
\begin{table}[t]
\begin{center}
\begin{tabular}{|c|c|c|c|c|c|c|c|}
	\hline
	CLS	&	$ax$			&	$\Gamma/|\vec{q}_\pi|$ &	$g_{\Bscal B\pi}~[\GeV]$	&$a\delta$ &$a\Delta$	&	$h$	&	$\widetilde{g}$\\
	\hline
	B6 &	$-0.0156(4)$ &	$0.92(4)$ & $27.4(0.1)(0.6)$ &	0.141(4) &	 $0.034(4)$	&	$0.85(3)(2)$	&	$-0.122(7)$\\
    	E5 &	$-0.0238(9)$ &	$0.86(7)$ & $26.4(0.1)(1.0)$ &	0.133(6) &	 $-0.012(6)$	&	$0.82(3)(3)$ 	&	$-0.117(6)$\\  
	F6 &	$-0.0161(3)$ &	$0.95(3)$ & $27.7(0.1)(0.5)$ &	0.129(3) &	 $0.025(3)$	&	$0.86(3)(2)$ 	&	$-0.119(4)$\\  
	N6 &	$-0.0172(6)$ &	$0.88(6) $ & $26.6(0.1)(0.9)$ &	0.092(3) & $0.008(3)$ 	&	$0.82(3)(3)$	&	$-0.122(5)$\\
	\hline
 \end{tabular} 
 \end{center}
 \caption{ \label{tab:resultsh}Numerical values of $ax$, $\Gamma/|\vec{q}_\pi|$,  $g_{\Bscal B\pi}$, $\delta=m_{\Bscal} - m_B$, $\Delta=m_{\Bscal} - m_{B\pi}$, $h$ and $\widetilde{g}$ obtained on the four CLS ensembles that we have analyzed, with $m_{\Bscal}\approx m_B + m_\pi$.}
\end{table}
In the table, the first error on $h$ comes from the uncertainty on $m_{\Bscal}$ in the continuum limit and the second error comes from the error on $ax$. The light-quark mass and lattice spacing dependence is so small on our data that it is legitimate to try a fit with a constant: we obtain $h=0.84(3)$ and $\widetilde{g}=-0.120(3)$. Performing a linear fit in $m^2_\pi$, we get compatible results $h = 0.86(4)$ and $\widetilde{g}=-0.122(8)$. A third possibility is to use the NLO formulae of HM$\chi$PT \cite{DetmoldRB}
\begin{align}
h &= h_0\left[1-\frac{3}{4} \frac{3\widehat{g}^2_0 + 3 \widetilde{g}^2_0 + 2 \widehat{g}_0\widetilde{g}_0}{(4\pi f_\pi)^2} m^2_\pi \log m^2_\pi \right] + C_h m^2_\pi  \,, \label{fith:NLO1}\\
\widetilde{g}  &= \widetilde{g}_0  \left[   1 -  \frac{ 2+4 \widetilde{g}^2_0 }{(4\pi f_{\pi})^2}   m_{\pi}^2 \log(m_{\pi}^2)  \right] + C_{\widetilde{g}} m_{\pi}^2  \,, \label{fitg:NLO1}
\end{align}
where $\widehat{g}_0 = 0.5(1)$ \cite{OhkiPY,BulavaEJ} is the pionic couplings associated to $H^* \to H \pi$. We get $h=0.84(3)$ and $\widetilde{g} = -0.116(7)$. The previous formulae for $\widetilde{g}$ take into account corrections from tadpole diagrams to the axial coupling between
$J=0$ and $J=1$ heavy-light mesons while eq.~(\ref{fith:NLO1}) for $h$ is obtained by considering directly the strong vertex $H_{J_1} \to H_{J_2} \pi$ 
\cite{FajferHI}.
The quark mass dependence is very small and the influence of the chiral logarithms does not change our result significantly. We quote finally
\beq\label{hresult} 
h = 0.84(3)(2) \,, \quad \widetilde{g} = -0.122(8)(6) \,,
\eeq
where the first error is statistical and the second error corresponds to the uncertainty that we evaluate from the discrepancy between the constant and linear fits. We show in Figure~\ref{fig:chih} the chiral extrapolations of $h$ and $\widetilde{g}$.
Rigorously, in the NLO chiral fits, we have neglected the contribution from the heavy-light states of opposite parity, as computed in \cite{FajferHI}; they have been studied in \cite{BecirevicZZA}. Neglecting them is equivalent to assume $m_\pi \ll \delta=m_{\Bscal} - m_B$. 
Since, for our lattice ensembles, the pion mass lies in the range [280 -- 440]~MeV and the mass difference between the scalar $B$ meson and the ground state $B$ meson is of the order of $\delta \sim 400$~MeV, the contribution is not negligible. Therefore, we also tried the other fit formulae
\begin{align}
h=& h_0\left[1 -\frac{3}{4}\frac{3\widehat{g}^2_0 + 3\widetilde{g}^2_0 + 2\widehat{g}_0 \widetilde{g}_0}{(4\pi f_\pi)^2}  m^2_\pi \log(m^2_\pi) - \frac{h^2_0}{(4\pi f_\pi)^2}\frac{m^2_\pi}{2\delta^2}m^2_\pi \log(m^2_\pi) \right] + C^{\prime}_h m^2_\pi \,, \\
\widetilde{g}=& \widetilde{g}_0 \left[  1 -  \frac{ 2+4 \widetilde{g}^2_0 }{(4\pi f_{\pi})^2}   m_{\pi}^2 \log(m_{\pi}^2)  + \frac{ h^2_0}{(4\pi f_{\pi})^2} \frac{m_{\pi}^2}{8 \delta^2}  \left( 3 + \frac{\widehat{g}_0}{\widetilde{g}_0 } \right) m_{\pi}^2 \log(m_{\pi}^2)  \right] + C^{\prime}_{\widetilde{g}} m^2_\pi \,,
\end{align}
where the coupling $\widehat{g}_0$ is the same as before and the mass difference $\delta$ is given in Table \ref{tab:resultsh}. The results are $h = 0.85(3)$ and $\widetilde{g}=-0.116(7)$ and is also perfectly compatible with our previous findings. 

 \begin{figure}[t!]
	\centering 
	\includegraphics*[width=8.0cm]{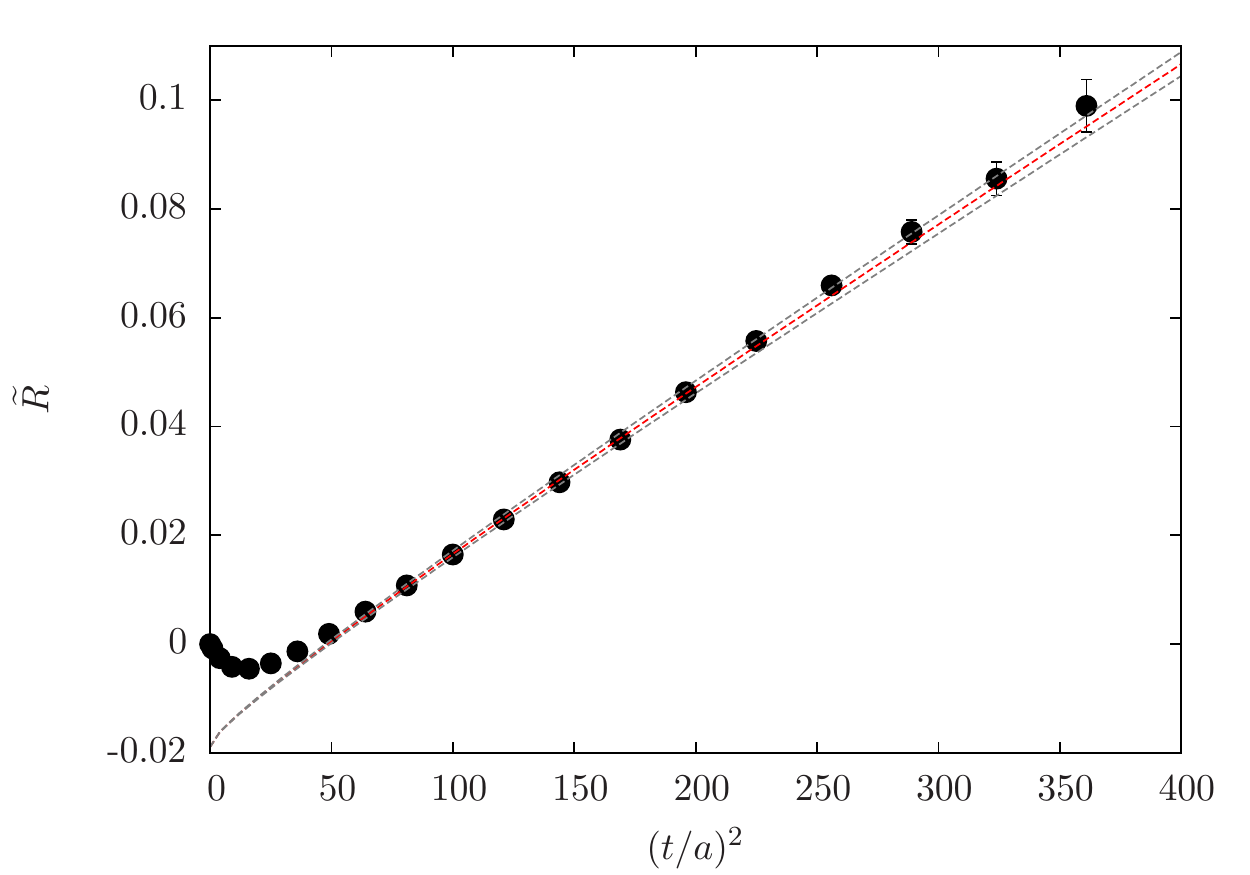}
	\caption{Quadratic fit of $\widetilde{R}(t)$ for the CLS ensemble E5.}
	\label{fig:Rtilde}
\end{figure}
In Refs.~\cite{McNeileXX,McNeileFH}, an alternative method to evaluate such a coupling like $h$ was proposed. Indeed, one can show that the connected contribution to the correlation function $C_{B \pi\; B \pi}(t)$, which includes box (\ref{BpiBpi2}) and cross (\ref{BpiBpi3}) diagrams, has the following behavior:
\beq\label{eq:Rtilde}
\widetilde{R}(t) =\frac{(v_{B\pi}(t),C_{\rm connected}(t) v_{B\pi}(t))}
{(v_{B\pi}(t),C_{B\pi\,B\pi}(t) v_{B\pi}(t))}= A^{\prime} +\frac{1}{2}x^2t^2 + {\cal O}(t) \,,
\eeq
where $C_{\rm connected}(t) = -\frac{3}{2}C_{\rm box}(t) +\frac{1}{2} C_{\rm cross}(t)$.
As explained before, these diagrams have been computed only for the CLS ensemble E5 and the function $\widetilde{R}(t)$ is plotted in Figure~\ref{fig:Rtilde}. The results are quite precise and the linear dependence in (\ref{eq:Rtilde}) cannot be neglected. Taking this into account, the result reads $|ax| = 0.0237(8)$, in perfect agreement with the one obtained by the previous method (see Table \ref{tab:resultsh}). The fit range has been varied from $t/a \in [9 - 18]$ to $t/a \in [13 - 18]$ where the result is stable to estimate the error.

\section{Discussion and conclusion}\label{sec6}

\noindent The couplings $h$ and $\widetilde{g}$ were explicitly computed on the lattice in Ref~\cite{BecirevicZZA}. For $h$, two results are reported for the two different actions used there: $h = 0.69(2)(^{+11}_{-7})$ and $h=0.58(2)(^{+6}_{-2})$. They are lower than what we get but this difference might be explained by the larger quark masses simulated at that time: indeed the chiral extrapolation tends to lower the extrapolated value. Our result is also a bit larger than the QCD sum rules estimates: in Ref~\cite{ColangeloPH} the computation of $g_{\Bscal B \pi}$ gives $h= 0.56(28)$, while in Ref~\cite{AlievBU} it gives $h = 0.74(23)$.\\
\noindent We can compare our finding with experimental data in the $D$ sector, although the static approximation of HQET is expected to give only a rough estimate due to quite large $1/m_c$ corrections. For example, in the case of the $D$ meson decay constant, a heavy quark spin breaking effects larger than 20\% between $f_D$ and $f_{D^*}$ has been measured \cite{BecirevicTI}. With $m_{D^*_0} = 2318(29)$ MeV and $\Gamma_{D^*_0}=267(40)$ MeV \cite{BeringerZZ}, we obtain $\Gamma(D^*_0 \to D \pi)/|\vec{q}_\pi| = 0.68(11)$ and $h = 0.74(8)$, assuming that the branching ratio ${\cal B}(D^*_0 \to D\pi)$ is $\sim$~100\%. This result is smaller than the one obtained in this work but it is compatible within error bars. In \cite{MohlerNA} the phase shift of the $D \pi$
scattering state was computed on the lattice: relating the coupling $g_{D^*_0 D \pi}$ quoted in that paper to $h$, one finds that $h$ is around 1.\\ 
\noindent Referring to the Adler-Weissberger sum rule \cite{AdlerGE}  in the $B \pi$ system, in the $m_Q\to \infty$ and soft pion limits, $\sum_\delta |X_{B\delta}|^2 = 1$, where $\Gamma ({\cal I} \to {\cal F} \pi)=\frac{1}{2\pi f^2_\pi} \frac{|\vec{q}|^3}{2j_{\cal I}+1} |X_{{\cal I} \to {\cal F}}|^2$ \cite{BecirevicFR}, we have the bound $\widehat{g}^2+h^2<1$. With the lattice average $\widehat{g}=0.5(1)$ made with the results \cite{OhkiPY, BulavaEJ}, we obtain that the sum rule would be saturated at 95\% by the $B^*$ pole and the first orbital excitation. \\
We also confirm the finding of Ref~\cite{BecirevicZZA} where a small value of $\widetilde{g}$ was obtained. In particular this coupling for positive parity states is smaller than in the case of negative parity states $\widetilde{g} \ll g$.\\
\noindent In conclusion, we have extracted from lattice simulations with ${\rm N_f}=2$ dynamical quarks the couplings $h$ and $\widetilde{g}$ that parametrise the emission of a soft pion by a scalar $B$ meson. We have observed a very mild quark mass and cut-off 
dependence of our numbers and we quote $h=0.84(3)(2)$, $\widetilde{g}=-0.122(8)(6)$ as our estimate. If $\widetilde{g}$ is small, the large value of $h$ compared to $\widehat{g} \sim 0.5$ outlines the fact that some care is necessary to apply HM$\chi$PT for pion masses close to mass splitting
$m_{B^*_0} - m_B \sim 400$ MeV: $B$ meson orbital excitation degrees of freedom cannot be neglected in chiral loops.

\section*{Acknowledgements}

We thank Damir Becirevic and Chris Michael for valuable discussions.
We are grateful to CLS for making the gauge configurations used in this
work available to us. Computations of the
relevant correlation functions are made on GENCI/CINES,
under the Grants 2013-056806 and 2014-056806.
N.G. acknowledges support from STFC under the grand ST/J000434/1.

\end{document}